\documentclass[10pt,journal,compsoc]{IEEEtran}

\usepackage{algorithmic}
\usepackage{algorithm}

\usepackage{amsthm}
\usepackage{amsmath}
\usepackage[braket, qm]{qcircuit}
\usepackage{subfigure}
\usepackage{setspace}
\usepackage{threeparttable}
\usepackage{float}
\usepackage{graphicx}
\usepackage{multirow}
\usepackage{booktabs}
\usepackage{color}
\usepackage{listings}
\usepackage{array}
\usepackage[utf8]{inputenc}
\usepackage{diagbox}
\usepackage{multirow}
\usepackage{makecell}
\usepackage{url}
\usepackage{graphicx}
\usepackage{array}
\usepackage{booktabs}

\ifCLASSOPTIONcompsoc
\usepackage[nocompress]{cite}
\else
\usepackage{cite}
\fi

\begin{document}

\title{Noise-resistant adaptive Hamiltonian learning}

\author{Wenxuan Wang,~\IEEEmembership{Member,~IEEE,} 
        
\IEEEcompsocitemizethanks{\IEEEcompsocthanksitem Wenxuan Wang is with the School of Computer Science and Engineering, Central South Univerisity, China, Changsha, 410083. 
\IEEEcompsocthanksitem Emails: 234701002@csu.edu.cn
}
\thanks{Manuscript received April 19, 2005; revised August 26, 2015.}}

%
%

\markboth{Journal of \LaTeX\ Class Files,~Vol.~14, No.~8, August~2015}%
{Shell \MakeLowercase{\textit{et al.}}: Bare Demo of IEEEtran.cls for Computer Society Journals}
%



\IEEEtitleabstractindextext{%
\begin{abstract}
Mitigating and reducing noise influence is crucial for obtaining precise experimental results from noisy intermediate-scale quantum (NISQ) devices. In this work, an adaptive Hamiltonian learning (AHL) model for data analysis and quantum state simulation is proposed to overcome problems such as low efficiency and the noise influence of quantum machine learning algorithms. First, an adaptive parameterized quantum circuit with noise resistant ability is constructed by decomposing the unitary operators that include penalty Hamiltonian in the topological quantum system. Then, a noise-resistant quantum neural network (RQNN) based on AHL is developed, which improves the noise robustness of the quantum neural network by updating iterative parameters. Finally, the experiments on Paddle Quantum demonstrate that RQNN can simulate the mathematical function and get accurate classification results on NISQ devices. Compared with the quantum neural network, RQNN ensures high accuracy with the same non-linear discrete data classification under the impact of amplitude damping noise, with an accuracy of 98.00 $\%$. It provides new possibilities for solving practical issues on NISQ devices and also benefits in the resolution of increasingly complicated problems, which will expand the range of potential applications for quantum machine learning models in the future.
	
\end{abstract}

}

\maketitle

\IEEEdisplaynontitleabstractindextext

%
\IEEEpeerreviewmaketitle

\IEEEraisesectionheading{\section{Introduction}\label{sec:introduction}}
\IEEEPARstart{Q}{uantum} computation has attracted considerable attention from the scientific community because of its unique quantum properties and the benefits of offering exponential acceleration for classical computing\cite{ref.TPAMI-PHL}.
In 2023, IBM proposed a quantum computer equipped with 1,121 qubits, representing a significant advancement in quantum computing ability.
The rapid increase of quantum computing resources has contributed to the improvements in quantum algorithms designed for noisy intermediate-scale quantum (NISQ) devices, especially in quantum machine learning \cite{TPAMI-00-liu2023deepeit,TPAMI-00-QNN}.
For example, Google introduced three types of learning tasks with "quantum advantage" in 2022, playing a pivotal role in studying quantum machine learning \cite{ref.google}.
Xiao \emph{et al.} introduced a generalized negation model to design the fundamental principles of quantum machine learning \cite{PAMI-05-Q}.
Sim \emph{et al.} took advantage of adaptive parameterized quantum circuits (PQC) to improve the robustness of quantum machine learning models\cite{ref.Sim}.
Valenti \emph{et al.} established a quantum error correction method in Hamiltonian learning and improved its fault-tolerant ability\cite{ref.Valenti}.
Hamiltonian learning (HL), introduced by Webie \emph{et al.} in 2014, is a quantum machine learning model with a significant simulation ability\cite{ref.Wiebe2014HL}, which acquires Hamiltonian information in quantum systems by training quantum circuits.
Hamiltonian is a fundamental component of quantum system and represents the energy of system.
Quantum state tomography\cite{ref.HL.application01}, chemical analysis\cite{ref.HL.application02}, and molecular structure prediction\cite{ref.HL.application03} are three domains where HL plays a significant role.
Recently, many HL models have been improved by using quantum machine learning methods, which has increased the computational efficiency of HL and expanded its application.
Shi \emph{et al.} developed a parameterized Hamiltonian learning and applied it to the image segmentation in 2022\cite{ref.TPAMI-PHL}.
Wang \emph{et al.} created a hybrid quantum-classical Hamiltonian learning method by designing quantum circuits to simulate the Hamiltonian on photonic  quantum computers\cite{ref.HL.wang}.
A series of optimization methods have advanced the investigation of HL models.
However, the noise resistance ability of HL have not been exhaustively investigated.
In general, researchers have mostly concentrated on improving the structure of quantum circuits in HL, but have paid little attention to addressing the negative noise influences.
The absence of noise-resistant quantum circuits for HL may impede its development and performance on NISQ devices.
Therefore, developing suitable quantum circuits to improve noise resistant ability of HL has become an important task.

Fortunately, we find inspirations in PQC \cite{ref.PQC} that show an adaptive method for designing quantum circuits\cite{ref.protein}.
One potential method for achieving this task is to design quantum gates with penalty Hamiltonian in PQC\cite{ref.Sim}, which can effectively improve the learning ability of HL by training parameters.
Numerous quantum machine learning models based on PQC have successfully finished the practical assignments, including drug discovery\cite{ref.applicationPQC01}, image classification\cite{ref.applicationPQC02}, and boson sampling\cite{ref.Boson}.
In summary, this research focuses on developing an adaptive PQC that is appropriate for HL and improves the noise resistance ability of the HL model.

In this work, Adaptive Hamiltonian Learning (AHL) is designed to improve the noise robustness of HL by decomposing and training the parameters in quantum gates with penalty Hamiltonian.
Moreover, a noise-resistant quantum neural network (RQNN) based on AHL is proposed, which uses gradient optimization to update the parameters.
Finally, inspired by Nguyen quantum neural network\cite{ref.nguyenQNN} and an adaptive correlation learning proposed by Li \emph{et al.} \cite{TPAMI-01-Li}, we have designed experiments on Paddle Quantum to demonstrate the effectiveness of RQNN in simulating mathematical function and data classification and the RQNN can produce accurate results in experiments even when amplitude damping noise is present, which show that AHL has a lot of possible applications in data analysis and mathematical functional design.

The primary contributions of this work are concluded as follows.
\begin{itemize}
	\item AHL and its adaptive parameterized quantum circuit are introduced, which improve the noise resistance ability of HL in NISQ. 
	AHL has remarkable ability in simulating quantum states and calculating ground state energy as a quantum machine learning method.

	\item RQNN based on AHL is proposed for improving the noise resistant ability of quantum neural network by training the parameters in PQC.
    RQNN can address mathematical function simulation and data classification tasks using the gradient descent optimization method, which has promise for practical application in data analysis.
    
    \item Experiments are carried out on the mathematical function simulation and data classification, demonstrating that RQNN can not only reduce the impact of noise, but also achieve high classification accuracy on the same non-linear discrete data classification under the impact of amplitude damping noise, with an accuracy of 98.00 $\%$, which is higher than the quantum neural network \cite{ref.QNN02}.
	
\end{itemize}

The subsequent sections of this paper are structured in the following. 
Section 2 overviews the related works and fundamental principles. 
The description of AHL can be obtained in  Section 3. 
Section 4 introduces RQNN based on AHL. 
Section 5 shows the detailed analyses and discussions of experiments. 
Finally, in section 6, conclusion is drawn.

\section{Related work}
In this section, the preliminary of quantum machine learning will be introduced first, then Hamiltonian learning  will be reviewed. Finally, parameterized quantum circuit will be shown. 
The notations that relate to our work are summarized in Table. (\ref{tab:notations}).

\begin{table}[htbp]
	\centering\small
	\begin{threeparttable}
		\caption{Notations}
		\label{tab:notations}
		\begin{tabular} {p{30pt}<{\centering}p{150pt}<{\centering}}
			\toprule
			\textbf{Notation}& \textbf{Description}\\
			\hline
			$H$ &  Hamiltonian \\
			$J$ &  Coupling Strength \\
			$U$ &  Unitary Operator \\
			$|\psi\rangle$ & Quantum State \\
			$R_{x}(\theta)$ & Quantum Rotation $X$ Gate \\
			$R_{z}(\theta)$ & Quantum Rotation $Z$ Gate \\
			$\sigma^{x}$ &  Pauli $X$ Operator  \\
			$\sigma^{z}$ &  Pauli $Z$ Operator  \\
			\bottomrule
		\end{tabular}
		\small
	\end{threeparttable}
\end{table}

\subsection{Preliminary}

The fundamental physical preliminaries of AHL and RQNN are described as follows.
	\setlength{\parindent}{0cm}

	\textbf{Quantum Gate:} 
    Quantum gate is the fundamental operating unit in quantum computing, comparable to the classical logic gate.
    A summary of the quantum gates that are frequently used in quantum machine learning can be found in the Appendix.
	In quantum machine learning, quantum gates can be represented by a unitary operator $U(\theta)$ as shown in Eq. (\ref{Unitory_theta}) \cite{ref.unitory}.
	\begin{equation}\label{Unitory_theta}
		U(\theta)=e^{iH\theta}
	\end{equation}
	where $i$ is the imaginary unit, $H$ is the Hamiltonian and $\theta$ is the parameter.
    Quantum gates with parameters can be obtained by matrix decomposition of unitary operators.

	\textbf{Hamiltonian:} 
	Hamiltonian is a mathematical operator that represents the overall energy of a quantum system.
	The unitary operator $U(\theta)$ contains the Hamiltonian $H$, as shown in Eq. (\ref{Unitory_theta}) \cite{ref.Wiebe2014HL}.
	The process of quantum computation can be understood as the result of the system evolution with the Hamiltonian, where the initial state is transformed into the final state through a series of quantum gates.
	
	\textbf{Topological Quantum System:}
	Topological quantum system is characterized by topological properties, which means that its properties remain constant in the face of minor perturbations under some noise influence\cite{ref.topological}.
	The Hamiltonian is an essential part for describing the topological system and its evolution.
	Adiabatic evolution is a frequently used approach in topological system evolution.
	During the adiabatic evolution, the quantum system can transit from one topological phase to another while maintaining its topological features, which can increase the robustness of quantum system to reduce noise influence.
	
	\textbf{Adiabatic evolution:}
    Adiabatic evolution is an approach used to describe the temporal evolution of a quantum system, where the system remains in an energy eigenstate throughout the process.
    The Hamiltonian of the quantum system changes as time t takes in adiabatic evolution and this change is typically shown in the Eq. (\ref{Adiabatic evolution}) \cite{ref.Adiabatic}.
    \begin{equation}\label{Adiabatic evolution}
    	H(t)=(1-s) H_b+s H_p,
    \end{equation}
    \begin{equation}\label{St}
    	s(t)=\frac{t}{T},
    \end{equation}
    where $s$ is the evolution parameter that is related to time, $H_b$ and $H_p$ are the Hamiltonian of the initial and final states respectively and $T$ is the total evolution time.
    The system experiences a progressive transition from energy level $H_b$ to $H_p$ by adjusting the change of $s$.

	\subsection{Hamiltonian Learning}
	\begin{figure}[t]
		\centering
		\includegraphics[width=0.45\textwidth]{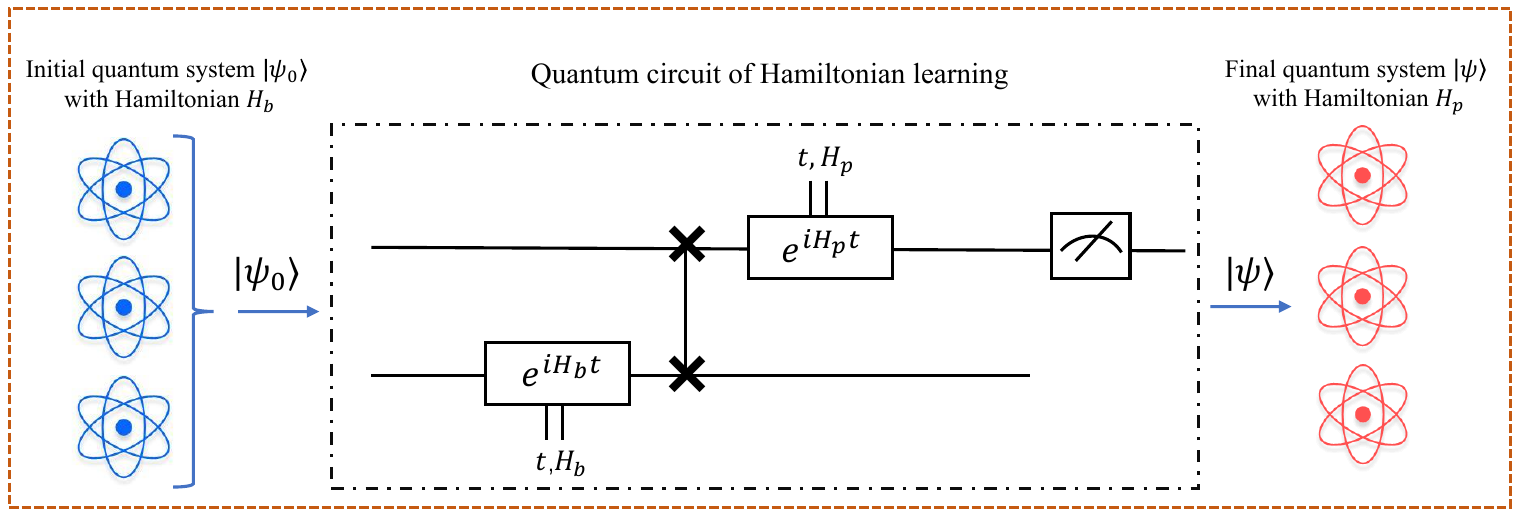}
		\caption{Hamiltonian learning model. The quantum state $|\psi_{0}\rangle$ is used to simulate the final quantum state with specific Hamiltonian $H_{p}$ by quantum system evolution.}
		\label{fig:Hamiltonian learning model 2014}
	\end{figure}
	
	\begin{figure*}[t]
		\centering
		\includegraphics[width=\textwidth]{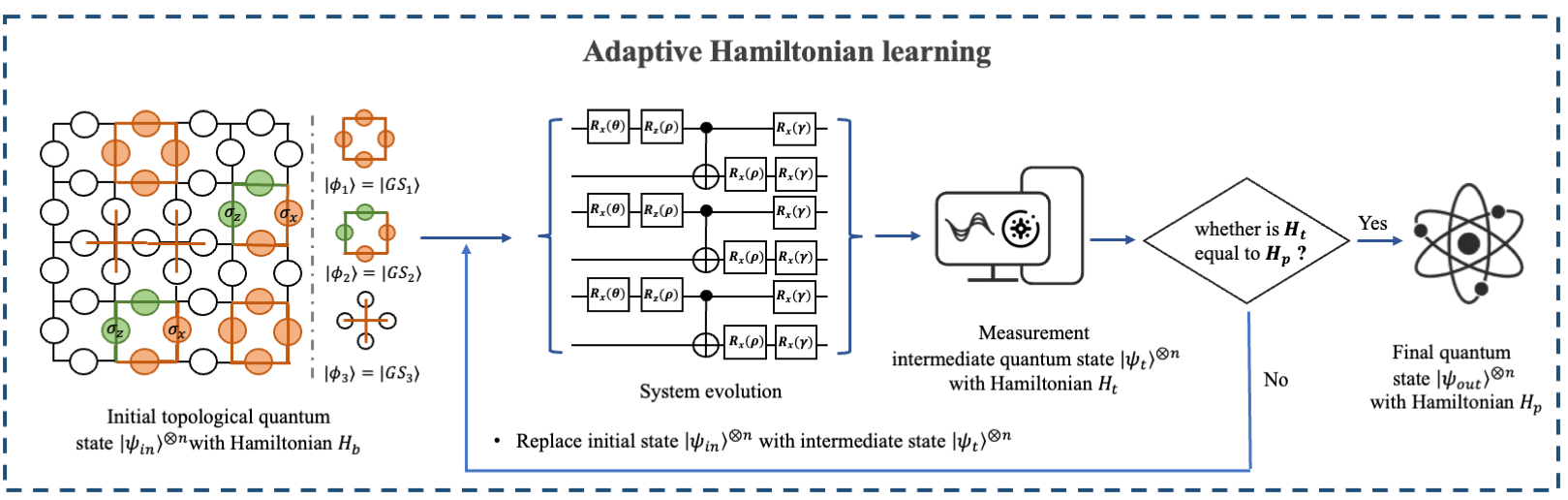}
		\caption{The adaptive Hamiltonian learning model.
			We design the topological quantum system $|\psi_{in}\rangle$ with Hamiltonian $H_{b}$ in the AHL as the starting quantum system. 
			The initial topological quantum system consists of Pauli $X$ operators (shown in orange) and Pauli $Z$ operators (shown in green) and there are three ground states $|GS_{t}\rangle, t \in {0,1,2}$.
			The quantum gates include Pauli rotation $X$ gates, Pauli rotation $Z$ gates and CNOT gates.
			The final quantum system $|\psi_{out}\rangle$ with Hamiltonian $H_{p}$ has adaptive ability to reduce the impact of noise. }
		\label{fig:AHL-model}
	\end{figure*}
	
	Hamiltonian learning (HL) is a quantum machine learning model for learning and simulating the Hamiltonian of a quantum system.
    Fig. (\ref{fig:Hamiltonian learning model 2014}) describes the process of HL, where the initial quantum system is used to simulate the final quantum system that has specific Hamiltonian.
	In essence, HL applies Bayes theorem to determine the probability that a given hypothesis about the Hamiltonian $H$ is accurate, which is based on the information gathered from experimentation\cite{ref.Wiebe2014HL}. 
	
	The implementation method of HL is as outlined below.
	First, an initial quantum state $|\psi_{0}\rangle$ with Hamiltonian $H_0$ needs to be created. 
    Then, the initial quantum system experiences a series of quantum gates to complete system evolution. 
    The quantum gates depicted in Fig. (\ref{fig:Hamiltonian learning model 2014}) are derived by the decomposition of the unitary operators about the Hamiltonian with time $t$.
	By permitting quantum circuits to complete quantum system evolution through swap gates, HL can satisfy the conditions of Bayes theorem in Eq. (\ref{Bayes}).
	\begin{equation}\label{Bayes}
		\operatorname{Pr}(H \!\mid\! D)\!=\!\frac{\operatorname{Pr}(D \mid H) \operatorname{Pr}(H)}{\operatorname{Pr}(D)}\!=\!\frac{\operatorname{Pr}(D \mid H) \operatorname{Pr}(H)}{\int \operatorname{Pr}(D \mid H) \operatorname{Pr}(H) \mathrm{d} H},
	\end{equation}
	where $H$ is a given hypothetical Hamiltonian, $D$ is the observed data and $\operatorname{Pr}$ is the prior and the likelihood function. 
	The Born rule provides the probability function for these likelihood functions in Hamiltonian $H$ as follows in Eq. (\ref{Born}).
	\begin{equation}\label{Born}
		\operatorname{Pr}(D \mid H)=\left|\left\langle D\left|e^{-i H t}\right| \psi\right\rangle\right|^2,
	\end{equation}
	where $|\psi\rangle$ is the quantum state.
    In the end, the quantum state information can be obtained by measurement.

	\subsection{Parameterized Quantum Circuit}
    Parameterized quantum circuit (PQC) is one of the core technologies in quantum machine learning \cite{ref.PQC}, which is based on a hybrid quantum-classical framework. 
    Through training and updating circuit parameters, it seeks to present the "quantum advantage" and accomplish quantum machine learning tasks such as classification, simulating mathematical function and picture recognition. 
    PQC comprises a series of quantum gates, including stationary and parameterized rotating gates. 
    Rotation angles of quantum gates with parameters are trained by classical optimizers.
    Quantum Neural Network (QNN) \cite{ref.QNN01,ref.QNN02} is a specific application of PQC in quantum machine learning. 
    It integrates the structures of classical neural networks with the properties of quantum mechanics to get the "speed advantages" of quantum computers and improve the computational efficiency for current computers. 
    Each neuron in the QNN can be represented by PQC.  
    Like classical neural networks, QNN has numerous layers and each layer contains multiple neurons. 
    The computation ability of QNN can be optimized by adjusting the weights of the parameters in PQC through training and updating.

\section{Adaptive Hamiltonian Learning}
In this section, we present an adaptive Hamiltonian Learning model (AHL) and its parameterized quantum circuit.
Initially, we provide the Hamiltonian and topological quantum system of AHL. 
Next, the construction of its parameterized quantum circuit will be demonstrated.

\subsection{AHL Modle}

\begin{figure}[bt]
	\[ \Qcircuit @C=1.3em @R=.9em {
		& \qw & \gate{R_{x}(\theta)}& \gate{R_{z}(\rho)}& \ctrl{1} & \qw & \gate{R_{x}(\gamma)} & \qw & \qw\\
		& \qw & \gate{R_{x}(\theta)}& \qw & \targ & \gate{R_{x}(\rho)} & \gate{R_{x}(\gamma)} & \qw & \qw \gategroup{1}{3}{2}{7}{1em}{--} \\
	} \]
	\caption{The parameterized quantum circuit of AHL.The circuit comprises Pauli rotation $X$ gates and Pauli rotation $Z$ gates. Each parameters need to be updated.}
	\label{fig:parameterized quantum circuit of AHL}
\end{figure}
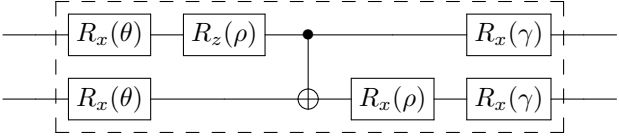

Fig. (\ref{fig:AHL-model}) depicts the framework of AHL, where an initial topological quantum system evolves adiabatically to a final state that can reduce noise influence adaptively.
Contrary to HL, we construct a quantum model with penalty Hamiltonian in Eq. (\ref{Hamiltonian}), which
plays an important role on AHL to reduce the impact of noise on the final quantum state.
Fig. (\ref{fig:AHL-model}) (a) presents an initial topological quantum system, where quantum information is stored in topological structures that are resistant to noise influences.
The initial quantum system of AHL is shown in a $k \times k$ square lattice of a topological quantum system  that features the periodic boundary conditions and the spin of freedom that situated along the edges is $-\frac{1}{2}$, $\frac{1}{2}$ degrees.
The quantum system comprises the Pauli $X$ operators depicted in orange and the Pauli $Z$ operators depicted in green, as specified in Eq.  (\ref{quantum_system_paoli_1}) and Eq. (\ref{quantum_system_paoli_2}). 
\begin{equation}\label{quantum_system_paoli_1}
	A_{n}=\sum_{n=1}^{N}\sigma_{n}^{x}
\end{equation}
\begin{equation}\label{quantum_system_paoli_2}
	B_{s}=\sum_{j=1}^{S}\sigma_{j}^{z},
\end{equation}
where $\sigma_{n}^{x}$ is the Pauli $X$ operator and $\sigma_{j}^{z}$ is the Pauli $Z$ operator, $n=1,2,3,...,N$ and $N$ is the number of Pauli $X$ operator, $j=1,2,3,...,S$ and $S$ is the number of Pauli $Z$ operator. 
Some Paoli $X$ operators are interconnected and entangled with Paoli $Z$ operators.
The ground state of the topological quantum system depends on the topology characteristic and Pauli operators. 
As shown in Fig. \ref{fig:AHL-model} (a), there are three kinds of ground states $|GS_{t}\rangle, t \in {0,1,2}$.
$|GS_{1}\rangle$ and $|GS_{3}\rangle$ are the ground states composed of Paoli X operators, where the spin angle of $|GS_{1}\rangle$ is $\frac{1}{2}$ and the spin angle of $|GS_{3}\rangle$ is $ -\frac{1}{2}$.
$|GS_{2}\rangle$ is the ground state with Paoli Z operators and Paoli X operators, and the spin angle is $\frac{1}{2}$. 
Eq. (\ref{Hamiltonian_initial}) is the Hamiltonian of the initial topological quantum system.
\begin{equation}\label{Hamiltonian_initial}
	\begin{aligned}
		H_{b}
		=\sum_{n=1}^{N}\pi V_{n}\sigma_{x}^{n},
	\end{aligned}
\end{equation}
where $\sigma_{x}$ is the Pauli $X$ operator, $V_{n}$ is the nuclear spin shift, $n=1,2,3,...,N$ and $N$ is the number of Pauli $X$ operators. 
The final Hamiltonian $H_{p}$ constructed by the Pauli $X$ operator and Pauli $Z$ operator as shown in Eq. (\ref{Hamiltonian}),
\begin{equation}\label{Hamiltonian}
	\begin{aligned}
		H_{p}=\sum_{j=k=1}^{S}\frac{\pi}{2}J_{jk}(\sigma_{z}^{j}+\sigma_{x}^{k})+\sum_{n=1}^{N}\hbar\sigma_{x}^{n},
	\end{aligned}
\end{equation}
where $\sigma_{x}$ and $\sigma_{z}$ are respectively the Pauli $X$ operator and the Pauli $Z$ operator that compose the quantum system for AHL, $J_{jk}$ is the coupling strength between the j-th  Pauli $Z$ operator and the k-th Pauli $X$ operator, $n=1,2,3,...,N$ and $N$ is the number of Pauli $X$ operator in the quantum system, $j,k=1,2,3,...,S$ and $S$ is the number of Pauli $Z$ operator in the quantum system, $\hbar$ is the Planck constant. 
The first component in $H_{p}$ is the olap Hamiltonian $H_{olap}$ in Eq. (\ref{olap}).
\begin{equation}\label{olap}
	\begin{aligned}
		H_{olap}=\sum_{j=k=1}^{S}\frac{\pi}{2}J_{jk}(\sigma_{z}^{j}+\sigma_{x}^{k}),
	\end{aligned}
\end{equation}
which refers to the energy produced through the interplay of Pauli X operators and Pauli Z operators in the coupling and entanglement process of the quantum system.
The second component is the penalty Hamiltonian $H_{redun}$ in Eq. (\ref{redun}).
\begin{equation}\label{redun}
	\begin{aligned}
		H_{redun}=\sum_{n=1}^{N}\hbar\sigma_{x}^{n},
	\end{aligned}
\end{equation}
which aims to mitigate the impact of noise on the quantum system and improve robustness ability of AHL.
If noise interference occurs during the evolution process, the penalty Hamiltonian $H_{redun}$ in Eq. (\ref{redun}) will allow PQC to adaptively rectify the present quantum state to the prior quantum state.
It makes AHL method more appropriate for the operation mode of quantum devices in NISQ.
The topological quantum system with Hamiltonian $H_b$ experiences system evolution by PQC that includes $R_{x}(\theta)$, $R_{z}(\rho)$, $R_{x}(\rho)$, $R_{x}(\gamma)$, and CNOT gate. 
The result of system evolution is the intermediate state $|\psi_{t}\rangle $, which corresponds to the Hamiltonian $H_{t}$.
The evolution of the system occurs naturally and without interruption. 

\subsection{Parameterized Quantum Circuit of AHL}

The parameterized quantum circuit of AHL can establish quantum system with the Hamiltonian in Eq. (\ref{Hamiltonian_initial}) and Eq. (\ref{Hamiltonian}) and improve the noise resistant ability for HL models.
Quantum gates in parameterized quantum circuits can be represented by unitary operators $U$ in Eq. (\ref{unitary}).
The parameterized quantum circuit of AHL consists of the accumulation of unitary operators that are composed of parameters $\theta$, $\rho$, $\gamma$ and Hamiltonian $H_{b}$ and $H_{p}$.
\begin{equation}\label{unitary}
	\begin{aligned}
		&U(\theta,\rho,\gamma)\\
		&=\prod_{\alpha=1}^{L}U(H_{b},\theta_{\alpha})U(H_{p},\rho_{\alpha},\gamma_{\alpha})\\
		 &=\prod_{\alpha=1}^{L}U(H_{b},\theta_{\alpha})U(H_{olap},\rho_{\alpha})U(H_{redun},\gamma_{\alpha}) \\
		&=\prod_{\alpha=1}^{L}e^{-iH_{b}\theta_{\alpha}}\times e^{-iH_{olap}\rho_{\alpha}}\times e^{-iH_{redun}\gamma_{\alpha}},
	\end{aligned}
\end{equation}
where $\alpha=1,2,3,...,L$, $L$ is the number of training layers for parameterized quantum circuit of AHL, $\theta, \rho, \gamma$ are the parameters that need to be updated, $\theta_{\alpha}, \rho_{\alpha}, \gamma_{\alpha}$ are the parameters for AHL in $\alpha$-th layer, $i$ is the plural.
The parameterized quantum circuit of each layer is represented by the product of the unitary operators $U(H_{b},\theta_{\alpha})$, $U(H_{olap},\rho_{\alpha})$ and $U(H_{redun},\gamma_{\alpha})$, shown in Eq. (\ref{unitary01}), Eq. (\ref{unitary02}) and Eq. (\ref{unitary03}) respectively.
\begin{equation}\label{unitary01}
	U(H_{b},\theta_{\alpha})=e^{-iH_{b}\theta_{\alpha}},
\end{equation}
\begin{equation}\label{unitary02}
	U(H_{olap},\rho_{\alpha})=e^{-iH_{olap}\rho_{\alpha}},
\end{equation}
\begin{equation}\label{unitary03}
	U(H_{redun},\gamma_{\alpha})=e^{-iH_{redun}\gamma_{\alpha}}.
\end{equation}
By substituting the Hamiltonian from Eq. (\ref{Hamiltonian_initial}) and Eq. (\ref{Hamiltonian})  into Eq. (\ref{unitary}), we obtain the expression for the unitary operator $U(\theta, \rho, \gamma)$. 
The unitary operator $U(\theta, \rho, \gamma)$ includes the Hamiltonian and the parameters that need to be updated, as shown in equation (\ref{unitary04}). 
\begin{equation}\label{unitary04}
	\begin{aligned}
		&U(\theta,\rho,\gamma)=\prod_{\alpha=1}^{L}e^{-i\sum_{n=1}^{N}\pi V_{n}\sigma_{x}^{n}\theta_{\alpha}}\\
		&\times e^{-i\sum_{j=k=1}^{S}\frac{\pi}{2}J_{jk}(\sigma_{z}^{j}+\sigma_{x}^{k})\rho_{\alpha}}\times e^{-i\sum_{n=1}^{N}\hbar\sigma_{x}^{n}\gamma_{\alpha}},
	\end{aligned}
\end{equation}
By analyzing Eq. (\ref{unitary04}), it can be shown that the constituent quantum gates of AHL are $R_{x}$, $R_{z}$, and the $CNOT$ gate that represents the entanglement of qubits. 
The precise matrix decomposition for Eq.(\ref{unitary04}) can be found in Eq. (\ref{unitary05}).

    \begin{figure*}[bt] 
	\centering
	\begin{equation}\label{unitary05}
		\begin{aligned}
			&U(\theta,\rho,\gamma)=\prod_{\alpha=1}^{L}\left[e^{-i\sum_{n=1}^{N}\pi V_{n}\sigma_{x}^{n}\theta_{\alpha}}\times e^{-i\sum_{j=k=1}^{S}\frac{\pi}{2}J_{jk}(\sigma_{z}^{j}+\sigma_{x}^{k})\rho_{\alpha}}\times e^{-i\sum_{n=1}^{N}\hbar\sigma_{x}^{n}\gamma_{\alpha}}\right]\\
			&=\prod_{\alpha=1}^{L}\left[\prod_{n=1}^{N}e^{-i\pi V_{n}\sigma_{x}^{n}\theta_{\alpha}}\right]
			\times\prod_{\alpha=1}^{L}\left[\prod_{j=k=1}^{S}e^{-i\frac{\pi}{2}J_{jk}(\sigma_{z}^{j}+\sigma_{x}^{k})\rho_{\alpha}}\right]
			\times\prod_{\alpha=1}^{L}\left[\prod_{n=1}^{N}e^{-i\hbar\sigma_{x}^{n}\gamma_{\alpha}} \right]\\
			&=\prod_{\alpha=1}^{L}\left[\prod_{n=1}^{N}e^{-i\pi V_{n}\sigma_{x}^{n}\theta_{\alpha}}\right]
			\times\prod_{\alpha=1}^{L}\left[\prod_{j=k=1}^{S}e^{-i\frac{\pi}{2}J_{jk}\sigma_{z}^{j}\rho_{\alpha}}
			\times\prod_{j=k=1}^{S}e^{-i\frac{\pi}{2}J_{jk}\sigma_{x}^{k}\rho_{\alpha}} \right]
			\times\prod_{\alpha=1}^{L}\left[ \prod_{n=1}^{N}e^{-i\hbar\sigma_{x}^{n}\gamma_{\alpha}} \right]\\
			&=\prod_{\alpha=1}^{L}\left[\prod_{n=1}^{N}R_{x}(n,-\frac{1}{2\pi V_{n}}\theta_{\alpha}) \right]
			 \! \times \! \prod_{\alpha=1}^{L}\left[ \! \prod_{j=k=1}^{S}R_{z}(j,k,-\frac{1}{2\pi J_{jk}}\rho_{\alpha}) 
			 \! \times \! \!  \!  \prod_{j=k=1}^{S}R_{x}(j,k,-\frac{1}{2\pi J_{jk}}\rho_{\alpha})\right]
			 \! \times \! \prod_{\alpha=1}^{L}\left[ \! \prod_{n=1}^{N}R_{x}(n,-\frac{1}{2\hbar}\gamma_{\alpha}) \right],\\
		\end{aligned} 
	\end{equation}
	
\end{figure*}
In Eq. (\ref{unitary05}), the $R_{x}$ and $R_{z}$ are the Pauli rotation $X$ gates and Pauli rotation $Z$ gates respectively, $\sigma_{x}$ and $\sigma_{z}$ are the Pauli $X$ operator and the Pauli $Z$ operator for AHL, $V_{n}$ is the nuclear spin shift, $J_{jk}$ is the coupling strength between the j-th Pauli $Z$ operator and the k-th Pauli $X$ operator.
To establish a connection between the j-th Pauli Z operator and the k-th Pauli X operator, we use the CNOT gate. 
Fig. \ref{fig:parameterized quantum circuit of AHL} illustrates the parameterized quantum circuit based on AHL. 
It includes Pauli rotation $X$ gates, Pauli rotation $Z$ gates and CNOT gates.
Each distinct parameters in the parameterized quantum circuits of AHL will be updated.
After the initial quantum system $|\psi_{in}\rangle$ passes through the quantum gates in AHL, a final state quantum system with noise resistant ability $|\psi_{out}\rangle$ is obtained.

\begin{algorithm}[H]
	\caption{The Algorithm of RQNN.}\label{algorithm}
	\begin{algorithmic}
		
		\REQUIRE  $N_t$,  $N_s$, $N$, $M$, $R$, $E_t$, $y$, $x$, $H_b$ and $H_p$;
		\ENSURE loss function, accuracy of training set, the accuracy of the testing set.
		\STATE \textit{Step} 1: input and encode the classical dataset into an initial quantum state $|\psi_{in}\rangle$.
		\STATE \textit{Step} 2: use AHL model to build parameterized quantum circuits of RQNN neurons.
		\STATE \textit{Step} 3: complete the system evolution and get the generated and intermediate quantum system.
		\STATE \textit{Step} 4: optimize the measurement results and get the loss function, update the parameters in RQNN, and then retrain the neurons in RQNN. 
		\STATE For { $j<\xi$ :}
		\STATE
		\STATE \hspace{0.5cm}$\frac{\partial \mathcal{L}(\theta_{j}, \rho, \gamma)}{\partial \theta_{j}}=\frac{\mathcal{L}(\theta_{j}+\Delta_{j}, \rho, \gamma)-\mathcal{L}(\theta_{j}-\Delta_{j}, \rho, \gamma)}{2 \Delta_{j}}$;
		\STATE
		\STATE \hspace{0.5cm}$\theta_{j} \leftarrow \theta_{j}-R \frac{\partial \mathcal{L}(\theta_{j}, \rho, \gamma)}{\partial \theta_{j}}$;
		\STATE
		\STATE \hspace{0.5cm}$\frac{\partial \mathcal{L}(\theta, \rho_{j}, \gamma)}{\partial \rho_{j}}=\frac{\mathcal{L}(\theta, \rho_{j}+\Delta_{j}, \gamma)-\mathcal{L}(\theta, \rho_{j}-\Delta_{j}, \gamma)}{2 \Delta_{j}}$;
		\STATE
		\STATE \hspace{0.5cm}$\rho_{j} \leftarrow \rho_{j}-R \frac{\partial \mathcal{L}(\theta, \rho_{j}, \gamma)}{\partial \rho_{j}}$;
		\STATE
		\STATE \hspace{0.5cm}$\frac{\partial \mathcal{L}(\theta, \rho, \gamma_{j})}{\partial \gamma_{j}}=\frac{\mathcal{L}(\theta, \rho, \gamma_{j}+\Delta_{j})-\mathcal{L}(\theta, \rho, \gamma_{j}-\Delta_{j},)}{2 \Delta_{j}}$;
		\STATE
		\STATE \hspace{0.5cm}$\gamma_{j} \leftarrow \gamma_{j}-R \frac{\partial \mathcal{L}(\theta, \rho, \gamma_{j})}{\partial \gamma_{j}}$;
	\end{algorithmic}
\end{algorithm}

\section{Noise-resistant quantum neural network}

In this section, we present a noise-resistant quantum neural network (RQNN) that based on AHL. 
The RQNN uses a parameterized quantum circuit of AHL as its neuron unit, which possesses the ability to show exceptional noise robustness on NISQ devices. 
At first, we provide the framework of RQNN. 
And then, the algorithm of RQNN will be introduced.

\subsection{The framework of RQNN}

\begin{figure}[t]
	\centering
	\includegraphics[width=0.5\textwidth]{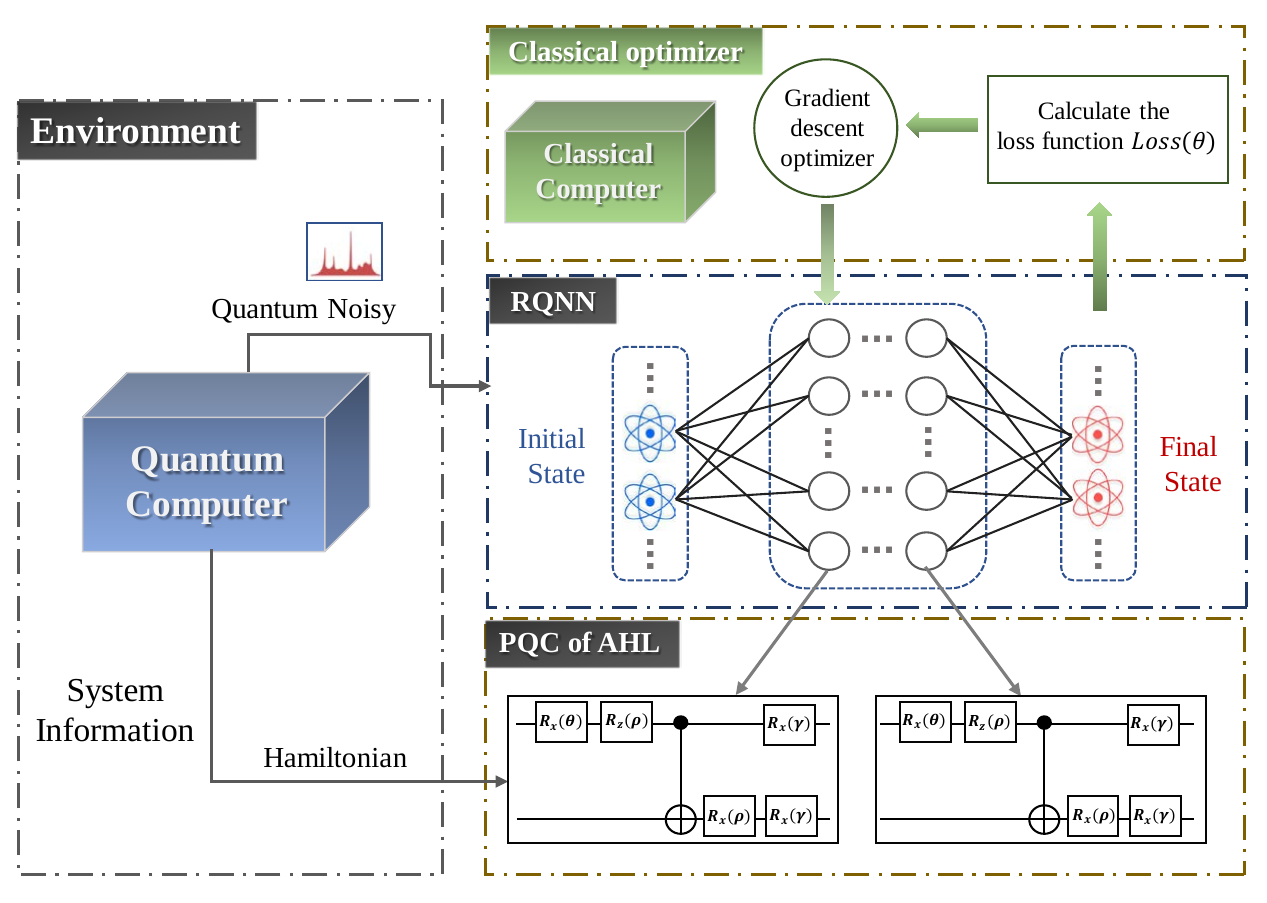}
	\caption{The network structure of RQNN.}
	\label{fig:RQNN-networkstructure}
\end{figure}

We present a RQNN based on AHL to update the parameters in neurons by using the gradient descent optimization approach, as shown in Fig.  (\ref{fig:RQNN-networkstructure}).
The initial quantum state $|\psi_{in}\rangle$ is the input vector in the RQNN, and the final quantum state $|\psi_{out}\rangle$ after the measurement is the output. 
The precise meaning of the initial quantum state $|\psi_{in}\rangle$ relates to the encoding method that is used to encode classical data into a quantum state.
The classical input data is encoded by the angle encoding approach, where the size of the input quantum state $|\psi_{in}\rangle$ consists of a series of unitary matrix vectors formed by quantum rotation gates.
In RQNN, each neuron is constructed by the parameterized quantum circuit in AHL, which means that the unitary matrix associated with the neurons in RQNN is created by decomposing the unitary operators with penalty Hamiltonian provided in AHL.
The approach combines the topological quantum system of adaptive noise resistant ability in AHL with the neural network structures in quantum machine learning models.
It improves the noise robustness ability of the existing quantum neural network.
The initial topological quantum system $|\psi_{in}\rangle$ experiences system evolution to an intermediate state $|\psi_{t}\rangle$,  which facilitated by the neurons in RQNN.
After evaluating the value of the Hamiltonian $H_{t}$ and the loss function, we use the gradient descent method to update the parameters in each neurons.
RQNN uses the  projection measurement with the z-axis \cite{ref.Boson} to map the final state $|\psi_{out}\rangle$ on the Z-axis of the Bloch sphere to get extracting classical information.
By analyzing and extracting the qubit information from the final quantum state $|\psi_{out}\rangle$ after measurement, RQNN can effectively perform quantum machine learning tasks, such as simulating mathematical function and data classification.
In summary, the RQNN is able to assess the output results by measuring the classical information in the final quantum state that has been evaluated and it has excellent noise resistant ability on NISQ devices.

\subsection{The algorithm for RQNN}

Algorithm. (\ref{algorithm}) presents the pseudocode of RQNN, which takes the following inputs: $N_t$, $N_s$, $g$, $N$, $M$, $R$, $E_t$, $y$, $x$, $H_b$, and $H_p$. 
We specify the values for the following parameters: the number of training sets $N_t$, the number of test sets $N_s$, the width of the decision boundary $g$, the number of qubits $N$, the depth of the quantum circuit $M$, the learning rate $R$, the number of training times $E_t$, the real data label $y$, and the data set $x$. 
$H_b$ is the initial Hamiltonian as shown in Eq. (\ref{Hamiltonian_initial}).
$H_p$ is the final Hamiltonian as shown in Eq. (\ref{Hamiltonian}).
The algorithm produces classical information after the measuring process.
The initial values of the parameters $\theta$, $\rho$, $\gamma$ in the neurons of RQNN are random.
After setting up the initial quantum environment for RQNN algorithm, we use the AHL method to build parameterized quantum circuits of RQNN neurons.
Then, the RQNN neurons can complete the system evolution and get the generated and intermediate quantum system.
Finally, RQNN optimizes the measurement results and calculates the loss function \cite{TPAMI-04-Bai,TPAMI-00-jiang2020co}.
The precise mathematical formula for the loss function of RQNN is demonstrated in Eq. (\ref{Eq.loss}) and Eq. (\ref{Eq.y}).
\begin{equation}
	\label{Eq.loss}
	L(\theta,\rho,\gamma)=\sum_{k=1}^W \frac{1}{W}\left|\tilde{y}^k-y^{k-1}\right|,
\end{equation}
\begin{equation}
	\label{Eq.y}
	y^{(i)}=(\theta,\rho,\gamma) x^{(i)}+\epsilon^{(i)},
\end{equation}
where $k = 1, 2, 3, ..., W$, $W$ is the total number of samples,$\tilde{y}$ is the final label value updated by RQNN, while $y$ represents the starting value label that is associated with a parameter.
The parameters $\theta, \rho, \gamma$ are variables that influence the behavior of the RQNN. 
The original data set is $x$, and $\epsilon$ represents the error, which decreases as the RQNN is trained. 
As the loss function reduces, the parameters $\theta$, $\rho$, and $\gamma$ in RQNN are updated according to Eq. (\ref{parameter01}), Eq. (\ref{parameter02}), and Eq. (\ref{parameter03}) \cite{TPAMI-05-Bai}.
\begin{equation}\label{parameter01}
	\theta_{j} \leftarrow \theta_{j}-R \frac{\partial \mathcal{L}(\theta_{j}, \rho, \gamma)}{\partial \theta_{j}},
\end{equation}
\begin{equation}\label{parameter02}
	\rho_{j} \leftarrow \rho_{j}-R \frac{\partial \mathcal{L}(\theta, \rho_{j}, \gamma)}{\partial \rho_{j}},
\end{equation}
\begin{equation}\label{parameter03}
	\gamma_{j} \leftarrow \gamma_{j}-R \frac{\partial \mathcal{L}(\theta, \rho, \gamma_{j})}{\partial \gamma_{j}},
\end{equation}
where $\theta_{j}$, $\rho_{j}$ and $\gamma_{j}$ are the parameters of the j-th step respectively.  
The $R$ is the learning rate.

\section{Experimental Results and discussion}

Quantum computation has demonstrated significant potential in diverse domains, particularly in machine learning and artificial intelligence. 
However, the practical application of quantum machine learning is constrained by the unique characteristics of quantum system, including the instability of qubits and the interference caused by noise. 
RQNN, as a novel quantum neural network framework, addresses these limitations by offering an improved quantum circuit based on AHL. 
It excels at performing efficient calculations in quantum computing environments with noise influence, which is poised to find broader applications in the near future. 
We attempt to use RQNN to address prevalent data analysis challenges in the realm of artificial intelligence. 
Through extensive investigations, we discover that RQNN had exceptional experimental results in both simulating mathematical function and data classification. 
\begin{figure}[ht]
	
	\begin{minipage}{1\linewidth}
		\vspace{4pt}
		\[ \Qcircuit @C=1.3em @R=.9em {
			& \qw & \gate{R_{x}(\theta)}& \gate{R_{z}(\rho)}& \ctrl{1} & \qw & \gate{R_{x}(\gamma)} & \qw& \qw\\
			& \qw & \gate{R_{x}(\theta)}& \qw & \targ & \gate{R_{x}(\rho)} & \gate{R_{x}(\gamma)} & \qw & \qw \gategroup{1}{3}{2}{7}{1em}{--} \\
		} \]
		\centerline{(a) The Neuron of RQNN.}
	\end{minipage}
	\begin{minipage}{1\linewidth}
		\[ \Qcircuit @C=1.3em @R=.9em {
			&\\
			& \qw & \gate{R_{x}(\alpha)}& \gate{R_{z}(\theta)}& \ctrl{1} & \qw & \qw & \qw\\
			& \qw & \gate{R_{x}(\alpha)}& \qw & \targ & \gate{R_{x}(\theta)}     & \qw & \qw  \gategroup{2}{3}{3}{6}{1em}{--} \\
		} \]
		\centerline{(b) The Neuron of QNN.}
	\end{minipage}	
	
	\caption{The network structure of QNN and RQNN for simulating function. }
	\label{fig:Neuron-data simulation}
\end{figure}

In this section, we present experiments and discussions about RQNN and AHL. 
Initially, we introduce the experiments on simulating mathematical functions to demonstrate that RQNN offers a precise method for simulating data curves, while also efficiently mitigating the impact of noise interference. 
Next, the experiments on data classification is presented, showing that RQNN has robustness ability and effectively improves the accuracy of classifying discrete data. 
Finally, we will give some discussions about RQNN and AHL.

\subsection{ Simulate Mathematical function}

\begin{figure*}[tb]
	\begin{center}
		\subfigure{\includegraphics[scale=0.2]{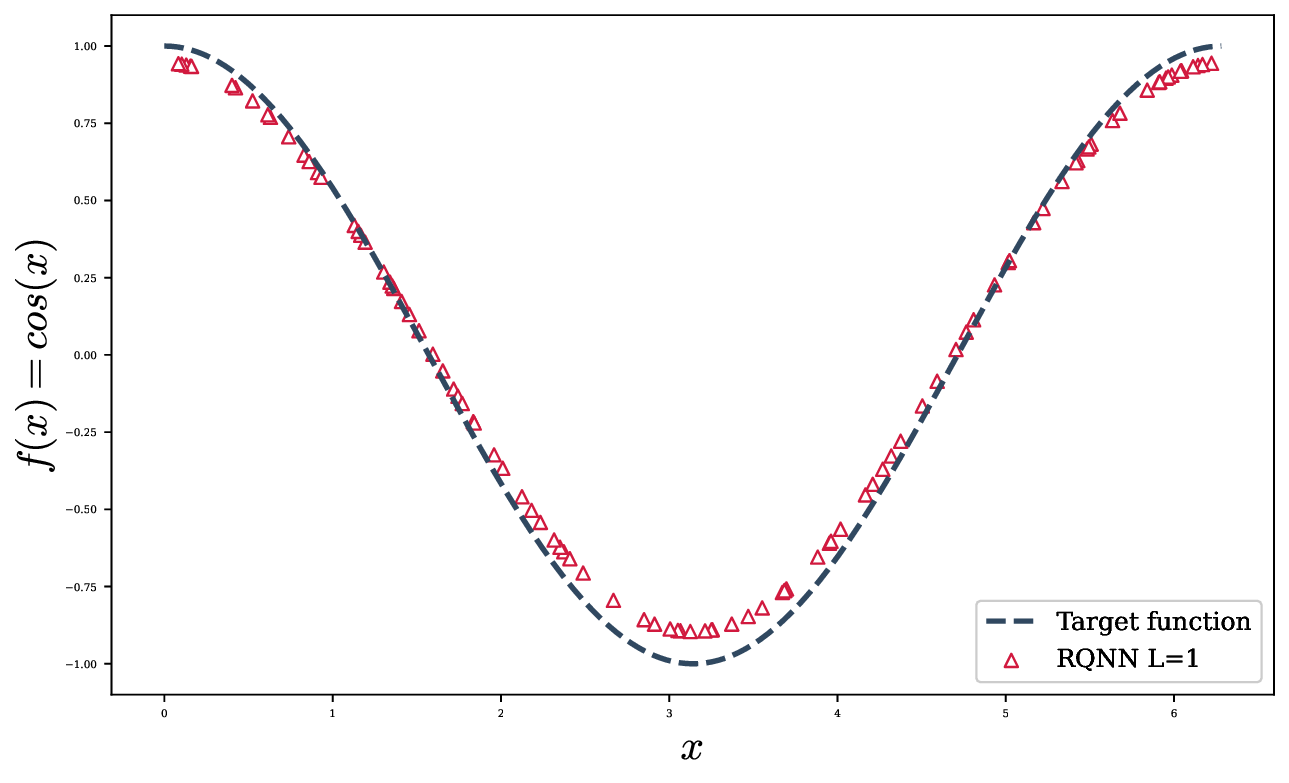}}
		\subfigure{\includegraphics[scale=0.2]{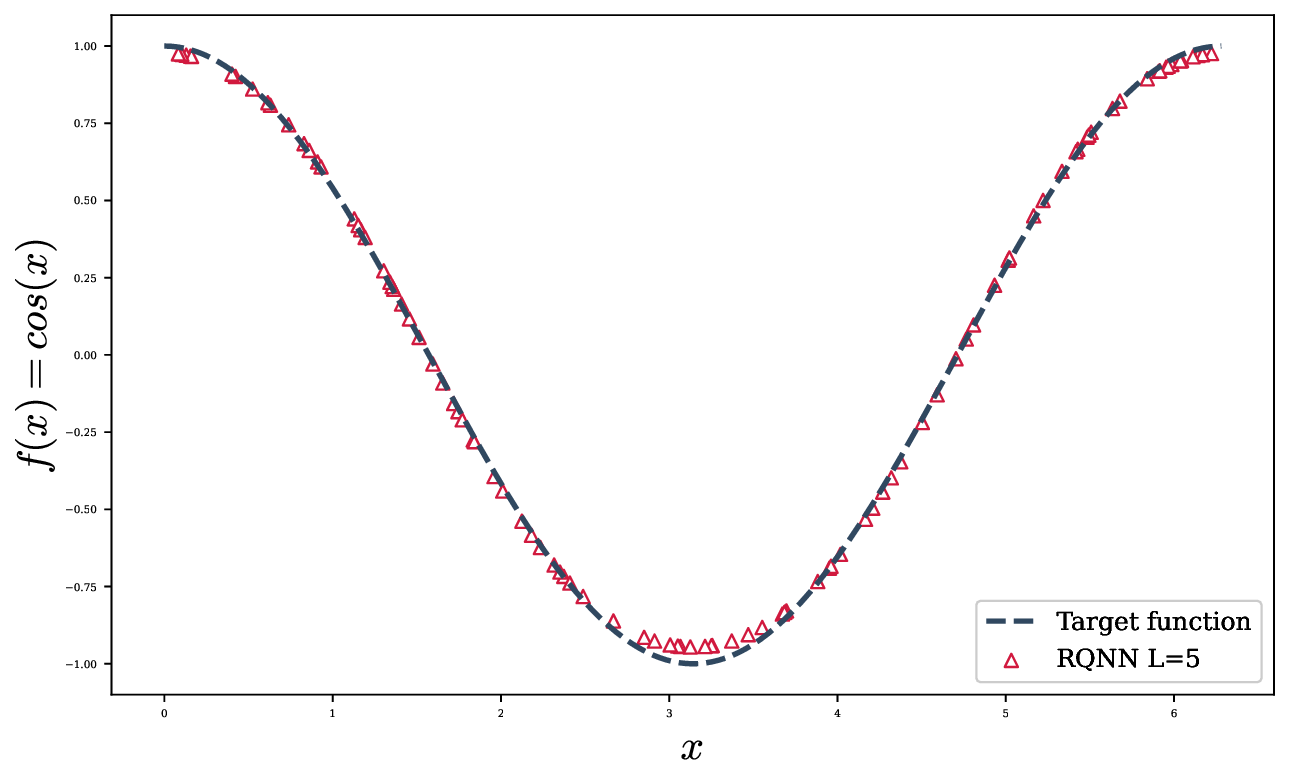}}
		\subfigure{\includegraphics[scale=0.2]{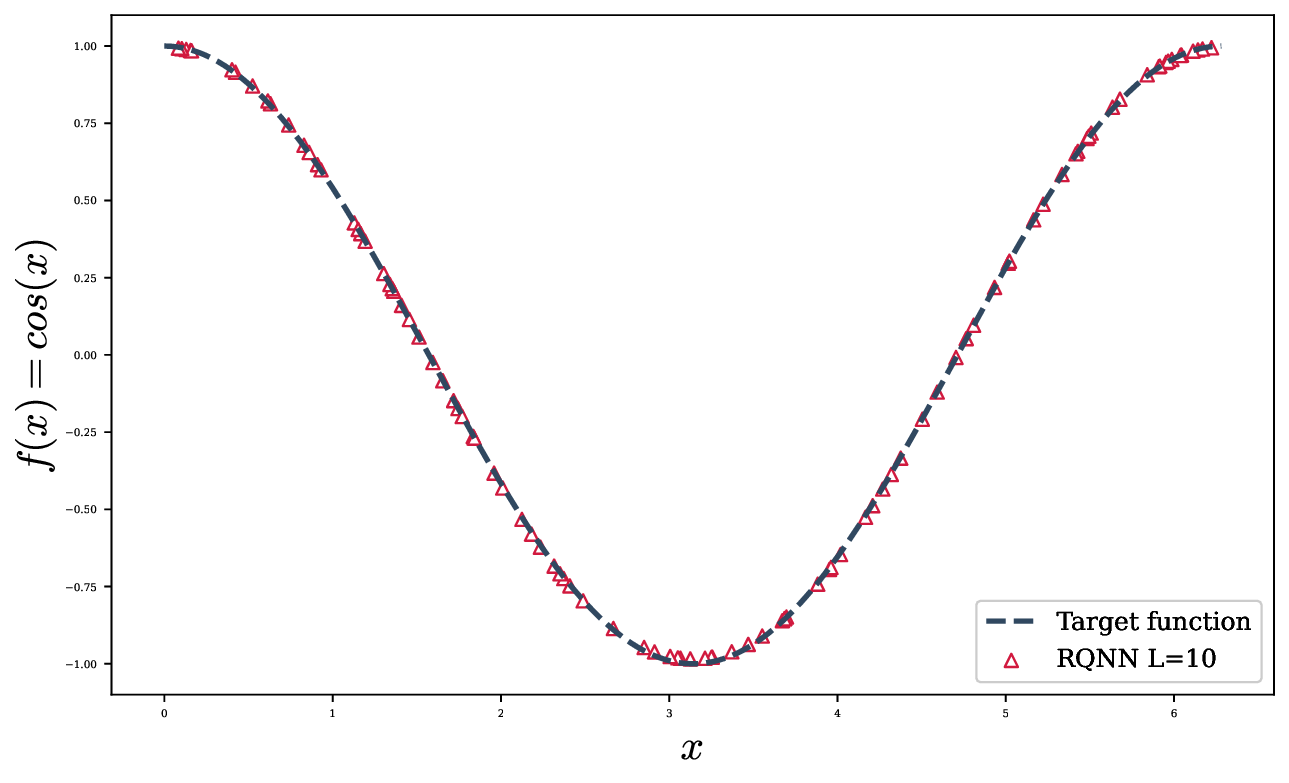}}
		\subfigure{\includegraphics[scale=0.2]{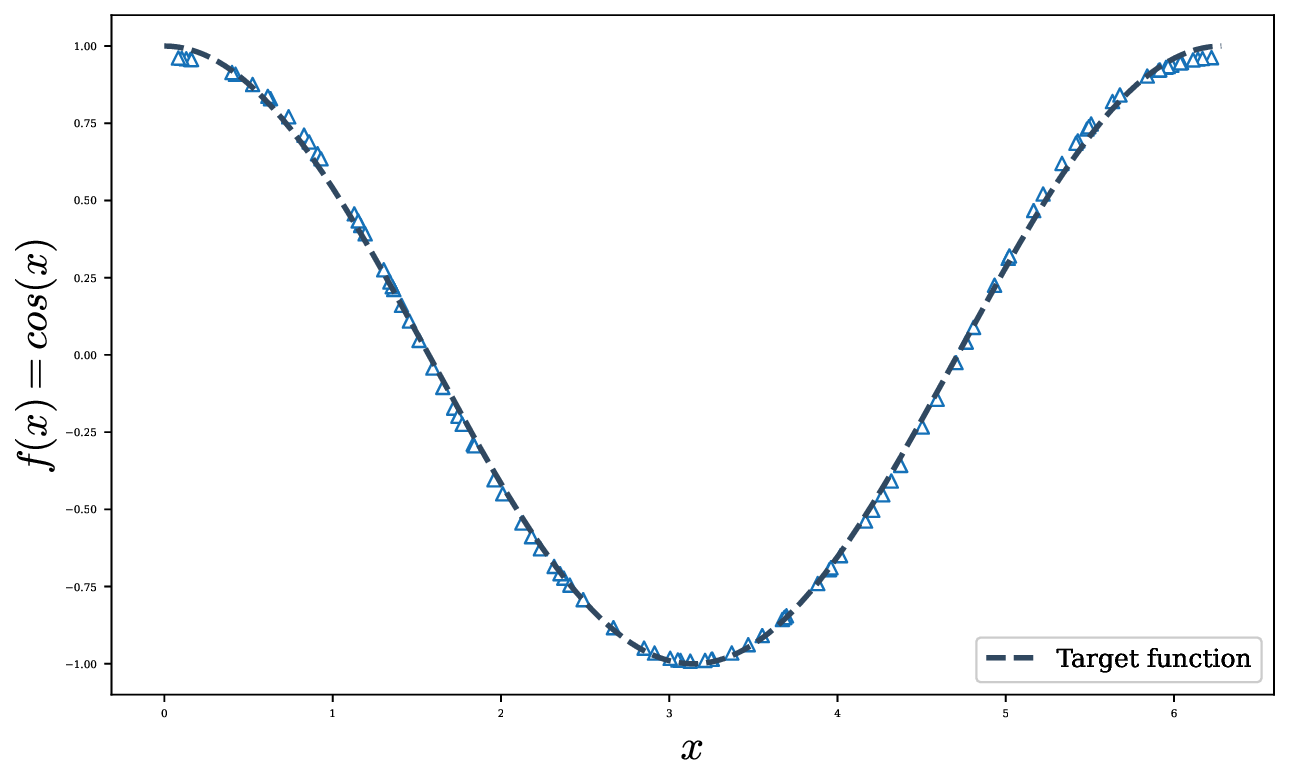}}
		
	\end{center}
	\caption{(a),(b),(c),(d) are the experiment results with RQNN for using data to simulate the function in $f(x)=cos(x)$. (a), (b), and (c) are the experiment results with different level for RQNN. (d) is the result for RQNN in Simulated noise-free quantum environment.}
	\label{fig:RQNN-COS}
\end{figure*}

\begin{figure*}[htbp]
	\centering
	\begin{minipage}{0.4\linewidth}
		\vspace{4pt}
		\centerline{\includegraphics[width=1\textwidth]{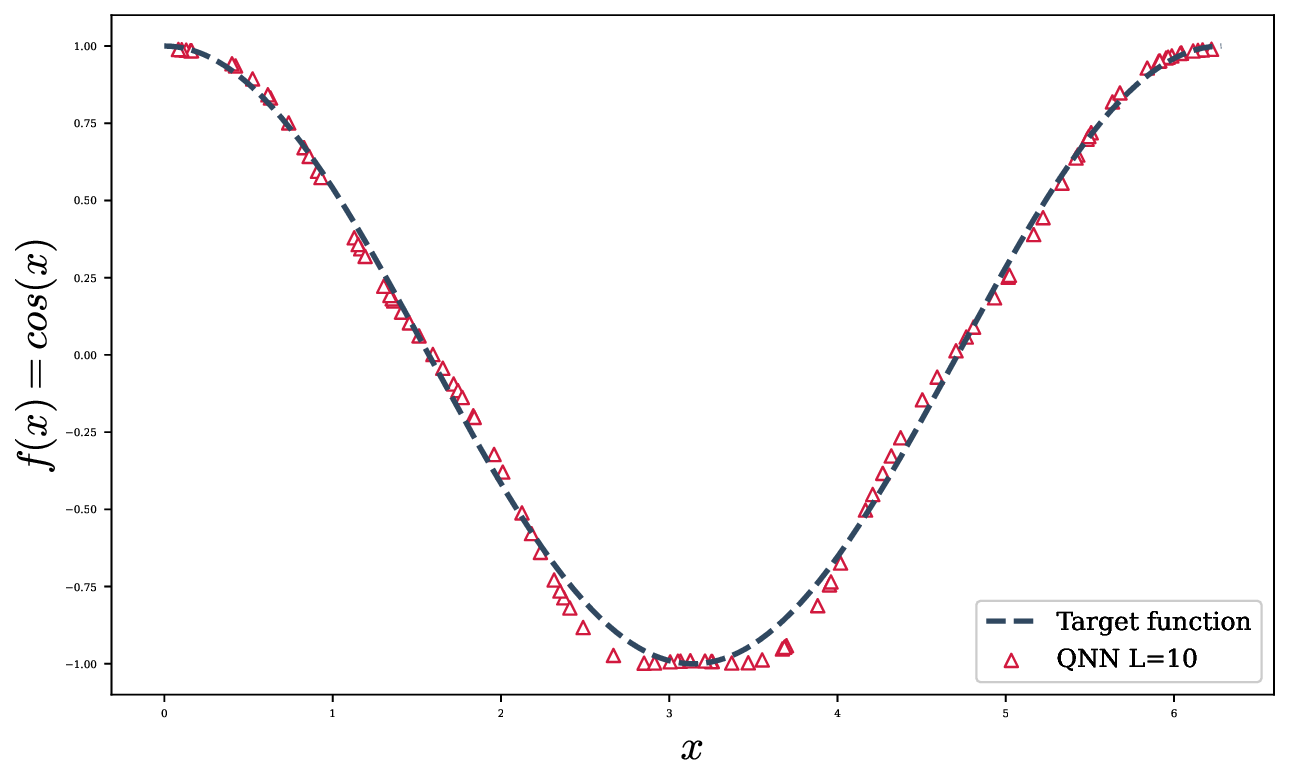}}
		\centerline{(a) The result of QNN.}
		
	\end{minipage}
	\begin{minipage}{0.4\linewidth}
		\vspace{4pt}
		\centerline{\includegraphics[width=1\textwidth]{RQNN-COS-10.eps}}
		\centerline{(b) The result of RQNN.}
	\end{minipage}
	
	\caption{The Experiment results with RQNN and QNN for training data to the cosine function and the results for RQNN in unnoisy . }
	\label{fig:COS}
\end{figure*}

RQNN can achieve the training of parameterized quantum circuits and optimization of parameters in neural elements, enabling NISQ devices to accurately represent mathematical function curves. 
To demonstrate the performance of RQNN in the area of noise resistant ability, we conduct a comparative experimental analysis between RQNN and quantum neural network (QNN) that does not have the penalty Hamiltonian. 
Fig. (\ref{fig:Neuron-data simulation}) (b) illustrates the neuron structure of QNN that includes quantum rotation X gate, quantum rotation Y gate, and CNOT gate.
To demonstrate the ability of RQNN in mitigating noise influence on NISQ devices and accurately performing artificial intelligence tasks, we introduce amplitude damping noise in the experimental quantum environment. 
Amplitude damping noise is a prevalent quantum noise commonly found in superconducting quantum computers.
we devised the following three sets of experimental operations.
\begin{enumerate}
	\item \textbf{Experiment (01)} : RQNN with different circuit depths to simulate the cosine function.
	\item \textbf{Experiment (02)} : RQNN and QNN with $10$ circuit depth to simulate the cosine function.
	\item \textbf{Experiment (03)} : RQNN and QNN with $10$ circuit depth to simulate the damped sine wave function.
\end{enumerate} 
The purpose of Experiment (01) is to investigate the influence of circuit depth on the training of RQNN parameters and to demonstrate the noise resistant ability of RQNN.
Experiments (02) and Experiment (03) are control experiments. 
Through the comparison of experiment results at equal quantum circuit depths, it can demonstrate that RQNN possesses both noise robustness and the ability to accurately simulate data for mathematical functions.

Fig. (\ref{fig:RQNN-COS}) shows the results from the test set of Experiment (01). 
Experiment (01) uses random sampling to generate data points for the cosine function. 
The training set consists of 600 discrete data points and the test set has 200 discrete data points. 
The learning rate of the RQNN is 0.2, and the model is trained for 300 iterations. 
As the circuit depth increases, the simulating accuracy of RQNN to the cosine function improves gradually. 
It will soon reach a level where it completely approximates the signal curve of the actual cosine function, which indicates that the ability of RQNN to resist noise also increases with the circuit depth. 
The effectiveness of RQNN is clearly shown when the quantum circuit is 10 depth. 
Fig. (\ref{fig:RQNN-COS}) (d) displays the results of simulating discrete data to the cosine function by training RQNN in a simulated environment without quantum noise. 
By comparing Fig. (\ref{fig:RQNN-COS}) (c) with Fig. (\ref{fig:RQNN-COS}) (d), we demonstrate the exceptional noise resistant ability of RQNN.

Fig. (\ref{fig:COS}) shows the results from the test set of Experiment (02), which simulate the cosine function on the test set using two distinct quantum neural network models with $10$ circuit depth. 
To mitigate the influence of extraneous variables on the experimental results, we preserved the pertinent parameter configurations of Experiment (01). 
By comparing Fig. (\ref{fig:COS}) (a) and Fig. (\ref{fig:COS}) (b), it can be shown that RQNN can efficiently and precisely simulate the data curve of the cosine function. 
But the experimental results of QNN deviate from the expected cosine function curve because of the impact of amplitude-damping noise. 
The experimental observation further supports the notion that the network structure of RQNN has noise resistant ability, and the AHL approach can significantly improve the stability and noise robustness of parameterized quantum circuits.
\begin{figure*}[htbp]
	\centering
	\begin{minipage}{0.4\linewidth}
		\vspace{4pt}
		\centerline{\includegraphics[width=1\textwidth]{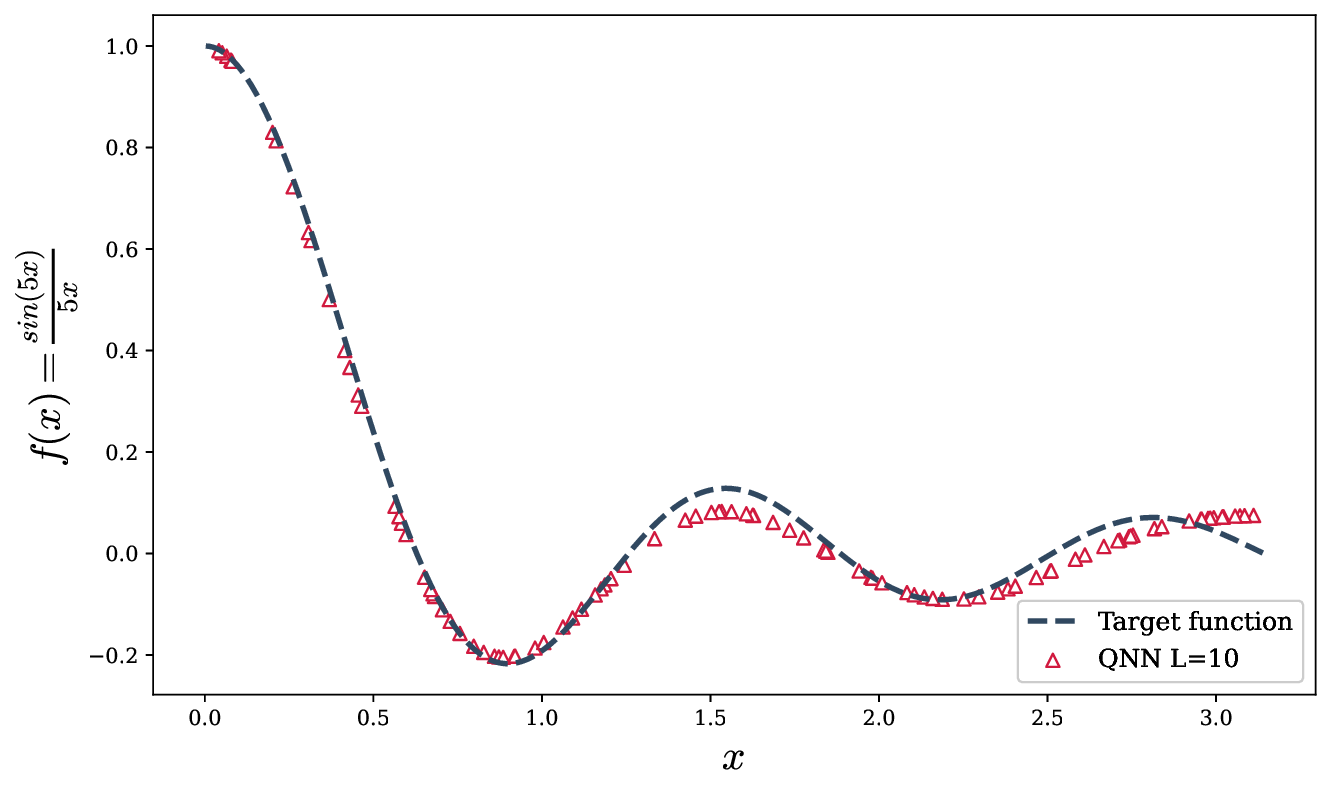}}
		\centerline{(a) The result of QNN.}
		
	\end{minipage}
	\begin{minipage}{0.4\linewidth}
		\vspace{4pt}
		\centerline{\includegraphics[width=1\textwidth]{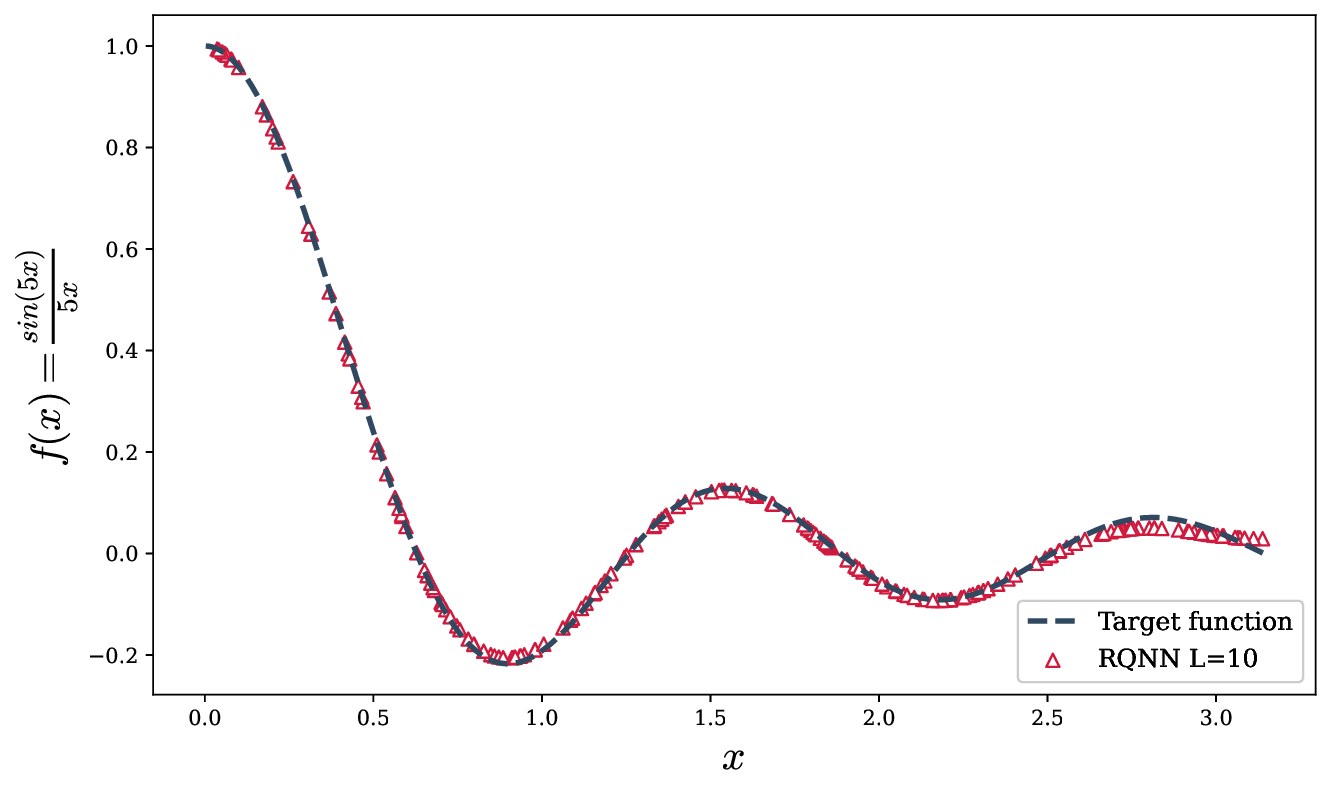}}
		\centerline{(b) The result of RQNN.}
	\end{minipage}	
	
	\caption{The Experiment results with RQNN and QNN for training data to simulate the damped sine wave function.}
	\label{fig:RQNN-QNN-Z}
\end{figure*}

Fig. (\ref{fig:RQNN-QNN-Z}) shows the results from the test set of Experiment (03),  which simulate the damped sine wave function on the test set using two distinct quantum neural network models with $10$ circuit depth.
Experiment (03) keeps the settings of irrelevant variables consistent with Experiment (01). 
The damped sine wave function has periodic behavior, with its graph displaying a steady attenuation of oscillation, which is used in physics, engineering, and biology. 
Through comparing Fig. (\ref{fig:RQNN-QNN-Z}) (a) and Fig. (\ref{fig:RQNN-QNN-Z}) (b) under the impact of amplitude damping noise, it is evident that there is a significant difference between the function simulating result of QNN and the actual function curve. 
But RQNN can accurately simulate the mathematical function when the quantum circuit depth is 10.
The precise simulation of the damped sine wave function demonstrates the substantial improvement in the robustness of RQNN compared to the QNN model.
Moreover, RQNN can simulate intricate mathematical functions accurately, making it extensively applicable in practical engineering. 

\begin{figure*}[htbp]
	\centering
	\includegraphics[width=0.9\textwidth]{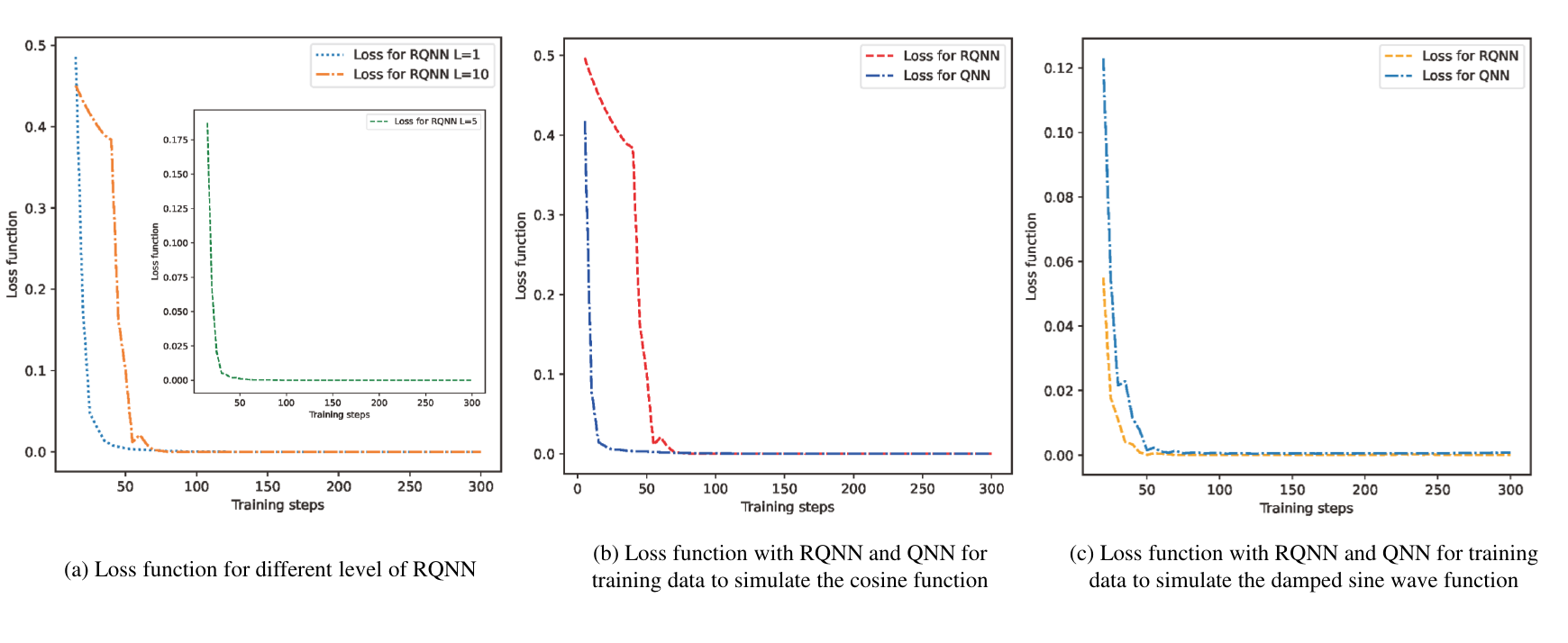}
	\caption{The loss function with RQNN and QNN.}
	\label{fig:Loss function}
\end{figure*}
Fig. (\ref{fig:Loss function}) displays the descent curve of the loss function. 
Precisely, Fig. (\ref{fig:Loss function}) (a) illustrates the results of the RQNN loss function of different circuit depths in Experiment (01). 
Fig. (\ref{fig:Loss function}) (b) and Fig. (\ref{fig:Loss function}) (c) introduce the loss function results of RQNN and QNN in Experiment (02) and Experiment (03) respectively. 
Based on the declining pattern of the curve, the loss function of RQNN can be rapidly minimized and converged. 
As the depth of the quantum circuit increases, the fluctuation of the loss function will also increase, which can be attributed to the rise in the number of parameters that need to be trained. 
The loss function of RQNN shows quick convergence, leading to iterative updates of the parameters in RQNN, which results in a more accurate function simulating effect and a significant improvement in the noise resistant ability of the quantum neural network.

\subsection{ Data Classification}

\begin{figure}[ht]
	
	\begin{minipage}{1\linewidth}
		\vspace{4pt}
		\[ \Qcircuit @C=1.3em @R=.9em {
			& \qw & \gate{R_{x}(\alpha)}& \gate{R_{z}(\theta)}& \ctrl{1} & \qw & \gate{R_{x}(\beta)} & \qw& \qw\\
			& \qw & \gate{R_{x}(\alpha)}& \qw & \targ & \gate{R_{x}(\theta)} & \gate{R_{x}(\beta)} & \qw & \qw \gategroup{1}{3}{2}{7}{1em}{--} \\
		} \]
		\centerline{(a) The Neuron of RQNN.}
	\end{minipage}
	\begin{minipage}{1\linewidth}
		\[ \Qcircuit @C=1.3em @R=.9em {
			&\\
			& \qw & \gate{R_{x}(\alpha)}& \qw& \ctrl{1} & \qw & \ctrl{1} & \qw& \qw\\
			& \qw & \gate{R_{x}(\alpha)}& \qw & \targ & \gate{R_{y}(\theta)}     & \targ & \qw & \qw \gategroup{2}{3}{3}{7}{1em}{--} \\
		} \]
		\centerline{(b) The Neuron of QNN.}
	\end{minipage}	
	
	\caption{The network structure of QNN and RQNN for data classification. }
	\label{fig:Neuron-for data classification}
\end{figure}

\begin{table}[ht]\footnotesize
	\setlength{\abovecaptionskip}{0cm}  
	\setlength{\belowcaptionskip}{-0.2cm} 
	\centering
	\caption{The experiment result of RQNN and QNN for data classification }
	\label{tab:Qubits results of Experiment 01}
	\setlength{\tabcolsep}{1.5mm}{
		\begin{tabular}{ c || c || c }
			\hline
			Model 	& Training dataset accuracy 	& Testing dataset accuracy 	   \\ 
			\hline
			QNN 	& $0.9000$		& $0.8333$	\\ 
			\hline 
			RQNN & $0.9800$ 		& $0.9867$ \\ 
			\hline 
			Gap & $0.0800$		& $0.1534$ \\ 
			\hline 
			
	\end{tabular}}
\end{table}

The quantum circuits of QNN \cite{ref.QNN02} and RQNN are shown in Fig. (\ref{fig:Neuron-for data classification}), for data classification experiment. 
The QNN that we choose is a prevalent structure in quantum classifier design, including quantum rotation gates and CNOT gates \cite{ref.QNN02}.
We use RQNN and QNN models to classify non-linear discrete data in Eq. (\ref{Eq.Data}) \cite{TPAMI-00hwang2011difference}.
\begin{equation}
	\label{Eq.Data}
	D=\left\{\left(x^k, y^k\right)\right\}_{k=1}^K,
\end{equation}
The overall dataset consists of 450 elements, with 300 elements in the training set and 150 elements in the test set. 
The decision boundary is 0.3, the learning rate is 0.1, and the models for RQNN and QNN  run 100 iterations.
The number of quantum qubits that are used is 4. 
The accuracy of data classification results is represented in Table. (\ref{tab:Qubits results of Experiment 01}), which reveals that the RQNN demonstrates a higher data classification accuracy on the test set compared to the QNN when subjected to amplitude damping noise. 
Moreover, the RQNN has higher accuracy ($98.00\%$) on the training set, which indicate that the RQNN effectively mitigates the impact of noise on the quantum machine learning models and significantly improves the ability to data analysis. 
Consequently, this experiment provides a scientific and theoretical foundation for using quantum machine learning like AHL in data processing applications.

\subsection{Discussion}

\begin{table*}[t]
	\centering
	\caption{The Comparison of Quantum Neural Networks that can mitigate noise influence}
	\label{tab:The Comparative Analysis of RQNN}
	\setlength{\tabcolsep}{0.8mm}{
		\begin{tabular}{ p{1cm}<{\centering}|p{4cm}<{\centering}|p{2cm}<{\centering}|p{2cm}<{\centering}|p{2cm}<{\centering}|p{1.5cm}<{\centering}|p{1.5cm}<{\centering} }
			\hline
			\multicolumn{2}{c|}{\textbf{Standard}} 
			&\multicolumn{1}{c|}{\textbf{Behrman \cite{ref.cp-RQNN-01}}}
			&\multicolumn{1}{c|}{\textbf{Krisnanda \cite{ref.cp-RQNN-02}}} 
			&\multicolumn{1}{c|}{\textbf{Patterson \cite{ref.cp-RQNN-03}}}
			&\multicolumn{1}{c|}{\textbf{Konar\cite{ref.cp-RQNN-04}}}
			&\textbf{RQNN}\\
			\hline
			\multicolumn{1}{c|}{\multirow{6}{*}{\textbf{General Character}}} 
			& \textbf{Mitigate Noise}
			& $\surd$
			& $\surd$
			& $\surd$
			& $\surd$
			& $\surd$ \\
			& \textbf{Fast Convergent Loss Function}
			& $\surd$
			& $\surd$
			& $\surd$
			& $\surd$
			& $\surd$ \\
			& \textbf{Updating Parameters}
			& $\surd$
			& $\surd$
			& $\surd$
			& $\surd$
			& $\surd$ \\
			& \textbf{Practical Application}
			& $\surd$
			& $\surd$
			& $\surd$
			& $\surd$
			& $\surd$  \\
			\hline
			\multicolumn{1}{ c|}{\multirow{6}{*}{\textbf{Individual Character}}} 
			& \textbf{Parameteried Quantum Circuit}
			& 
			& 
			& $\surd$ 
			& $\surd$ 
			& $\surd$\\
			& \textbf{Stable quantum circuit environment}
			& $\surd$ 
			& $\surd$
			& 
			& 
			& $\surd$ \\			
			& \textbf{Decomposable Unitory}
			& $\surd$ 
			&  
			& 
			& $\surd$ 
			& $\surd$ \\
			& \textbf{Adaptive quantum circuits}
			& 
			& 
			& 
			& 
			& $\surd$ \\
			\hline
			
	\end{tabular}}
\end{table*}

In Table. (\ref{tab:The Comparative Analysis of RQNN}), we conclude a comparison between RQNN and other quantum neural network models \cite{ref.cp-RQNN-01,ref.cp-RQNN-02,ref.cp-RQNN-03,ref.cp-RQNN-04} that are designed to reduce noise influence. 
The comparison standards are the algorithm design method and structure of the quantum neural network. 
The comparison results in Table. (\ref{tab:The Comparative Analysis of RQNN}) indicates that RQNN satisfies the standard for an anti-noise quantum neural network like its algorithm structure and quantum circuit design, which includes mitigating noise, fast convergent loss function, updating parameters and practical application. 
The most salient feature of the RQNN in quantum neural network architecture pertains to the design of adaptive quantum circuits and the training of parameters.

In Table. (\ref{tab:The Comparative Analysis of AHL}), we conclude a comparison between AHL and other Hamiltonian learning models \cite{ref.Wiebe2014HL,ref.Paesani,ref.Bairey,ref.HL.wang} that are used to simulate specific Hamiltonian quantum system. 
The comparison standards refer to the standards employed in quantum mechanics for assessing physical quantities and the progression of systems. 
It is evident that AHL meets the standards for overall character and offers the benefit of employing adaptive parameterized quantum circuits to mitigate the impact of noise.

The preceding experimental results clearly reveal that RQNN shows exceptional resistance to quantum noise and has outstanding skills in simulating mathematical function and data classification, which offers positive insights for the application of quantum machine learning.
Moreover, the experimental results further present the effectiveness of AHL in reducing noise influence on NISQ devices. 
Contrary to the quantum neural network, RQNN mitigates the impact of noise influence on NISQ devices, which provide a significant improvement in the precision of data analysis tasks. 
Based on the excellent performance on data analysis problems, it is possible to construct a quantum machine learning model like RQNN, which may facilitate the development of quantum machine learning in graph learning \cite{TPAMI-02-Li,TPAMI-03-Li}.

\section{Conclusion}
We establish the adaptive Hamiltonian learning (AHL) and investigate its ability to reduce noise influence on NISQ devices.
The proposed AHL presents a successful use of adaptive parameterized quantum circuits to improve the noise-resistance ability in quantum machine learning models, which establish a strong theoretical basis for future investigations. 
We design a RQNN based on AHL that has the ability to perform mathematical function simulation and data classification tasks. 
The experimental results illustrate that RQNN can improve the noise robustness ability of quantum neural networks and deliver precise simulation and classification results through updating the iterative parameters. 
We believe that with the development of NISQ devices, AHL can be incorporated with more quantum neural networks to address multi-dimensional data issue and have a broader variety of potential practical applications.

\begin{table*}[t]
	\centering
	\caption{The Comparison of Hamiltonian Learning Model}
	\label{tab:The Comparative Analysis of AHL}
	\setlength{\tabcolsep}{0.8mm}{
		\begin{tabular}{ p{1cm}<{\centering}|p{4cm}<{\centering}|p{2cm}<{\centering}|p{2cm}<{\centering}|p{2cm}<{\centering}|p{1.5cm}<{\centering}|p{1.5cm}<{\centering} }
			\hline
			\multicolumn{2}{c|}{\textbf{Standard}} 
			&\multicolumn{1}{c|}{\textbf{Wiebe \cite{ref.Wiebe2014HL}}}
			&\multicolumn{1}{c|}{\textbf{Paesani \cite{ref.Paesani}}} 
			&\multicolumn{1}{c|}{\textbf{Bairey \cite{ref.Bairey}}}
			&\multicolumn{1}{c|}{\textbf{Wang\cite{ref.HL.wang}}}
			&\textbf{AHL}\\
			\hline
			\multicolumn{1}{c|}{\multirow{6}{*}{\textbf{General Character}}} 
			& \textbf{Definition of Initial state }
			& $\surd$
			& $\surd$
			& $\surd$
			& $\surd$
			& $\surd$ \\
			& \textbf{Continuous System Evolution}
			& $\surd$
			& $\surd$
			& $\surd$
			& $\surd$
			& $\surd$ \\
			& \textbf{Preparation of Quantum State}
			& $\surd$
			& $\surd$
			& $\surd$
			& $\surd$
			& $\surd$ \\
			\hline
			\multicolumn{1}{ c|}{\multirow{6}{*}{\textbf{Individual Character}}} 
			& \textbf{Parameteried Quantum Circuit}
			& 
			& 
			&  
			& $\surd$ 
			& $\surd$\\
			& \textbf{Spontaneous System Evolution}
			& $\surd$ 
			& $\surd$
			& 
			& 
			& $\surd$ \\			
			& \textbf{Stable Adiabatic Process}
			& $\surd$ 
			& $\surd$
			& $\surd$ 
			& 
			& $\surd$ \\			
			& \textbf{Noise Resistant Ability}
			&  
			&  
			& 
			& 
			& $\surd$ \\
			\hline
			
	\end{tabular}}
\end{table*}

\bibliographystyle{IEEEtran}
\bibliography{IEEEfull,myref}


%

\vspace{11pt}

\vfill

\end{document}